\documentclass[showpacs,twocolumn,aps]{revtex4}
\usepackage{amssymb}
\usepackage{amsmath}
\usepackage{graphicx}
\usepackage{lscape}
\usepackage{booktabs}
\begin{document}
\title{$\rm ^3_{\Lambda}H$ and $\rm ^3_{\overline \Lambda}\overline H$ production
and characterization in Cu+Cu collisions at $\sqrt{s_{\rm{NN}}}=200$~GeV}
   \author{ Feng-Xian Liu$^{1,2}$, Gang Chen$^{1,2}$\footnote{Corresponding Author: chengang1@cug.edu.cn}, Zhi-Lei She$^{1,2}$, Liang Zheng$^2$, Yi-Long Xie$^2$, Zi-Jian Dong$^2$, Dai-Mei Zhou$^4$ and Ben-Hao Sa$^{3,4}$}

\address{
${^1}$Institute of Geophysics and Geomatics, China University of Geosciences, Wuhan 430074, China\\
${^2}$School of Mathematics and Physics, China University of Geosciences, Wuhan 430074, China\\
${^3}$ China Institute of Atomic Energy, P.O. Box 275 (10), Beijing 102413, China.\\
${^4}$ Institute of Particle Physics, Central China Normal University, Wuhan 430079, China.}

\begin{abstract}
Production of the (anti)hypertriton nuclei $\rm ^3_{\Lambda}H$ and $\rm ^3_{\overline \Lambda}\overline H$ in Cu+Cu interactions at $\sqrt{s_{\rm{NN}}}=200$~GeV is studied at three centrality bins of 0-10\%, 10-30\% and 30-60\% using the dynamically constrained phase-space coalescence model together with {\footnotesize{PACIAE}} model simulations. The $\rm ^3H(^3\overline H)$ and $\rm ^3He(^3\overline {He})$ nuclei has been compared. It is indicated by the study that the yields of light nuclei and hypernuclei increase rapidly from peripheral to central collisions while their anti-particle to particle ratios keep unchanged for different centrality bins. The strangeness population factor $S_3=\rm{{^3_{\Lambda}H}/({^3{He}}\times \Lambda/p)}$ was found to be close to unity, and was compatible with the STAR data and theoretical model calculation, suggesting that the phase-space population of strange quarks is similar with the ones of light quarks and the creation of deconfined high temperature quark matter in Cu+Cu collisions.

\end{abstract}
\maketitle

\section{Introduction}
The collisions of high-energy heavy-ion can recreate the condition similar
to that of the universe microseconds after the Big Bang and produce abundant nuclei and anti-nuclei in this process~\cite{star}. Thus, high-energy heavy-ion collisions provide us a unique opportunity to study and understand the production and property of nuclei and anti-nuclei under the extreme energy conditions.
Additionally, during the creation of deconfined quark gluon matter, the strange quark production is predicted to be largely enhanced due to thermal equilibrium, providing an important tool to produce light (anti-)hyper-nuclei~\cite{andr}.

Since Marian Danysz and Jerzy Pniewski made the first observation of the hyper-nucleus in a nuclear emulsion cosmic ray detector in 1952~\cite{dany}, it opened a new field, called hyper-nuclear physics, extending the research
scope of nuclear physics from the two-dimensional space of neutrons and protons
to the three-dimensional space including the hyperons. Hypernucleus contains
nucleon and hyperon ($\Lambda,\Xi,\Sigma,\Omega$). Hyperons are usually unstable particles with at least one strange quark in its valence components.

Hypertriton ($\rm ^3_{\Lambda}H$) is the lightest hypernucleus, which contains
a hyperon $\Lambda$, and two nucleons proton plus neutron. Since the first $\rm ^3_{\Lambda}H$ was observed,
several hypernuclei have been found, nevertheless, no anti-hypernucleus has ever
been discovery until the observation of the anti-hypertriton
($\rm ^3_{\overline \Lambda}\overline H$) by the STAR collaboration at Relativistic
Heavy-Ion Collider (RHIC) at Brookhaven National Laboratory (BNL) in 2010~\cite{abel}.
Hypernuclear physics is used to provide fascinating and fundamental
information about not only the interaction between the hyperon and the baryon(YN)
or the hyperon and the hyperon (YY), but also the influence caused by the hyperon
motion when it enters the nucleus. These interactions are related to nuclear
astrophysics and nuclear physics, such as the formation of the neutron stars~\cite{weber,heise,vida,lona}.
The productions of hyper-nuclei and several anti-hyper-nuclei have been studied
in heavy-ion collisions for many years, as mentioned earlier, for a review see Refs.~\cite{braun,e864,ma1,ma2,ma3,ma4,ma5}
and references therein.

 In this paper, the parton and hadron cascade model(PACIAE)~\cite{sa} is used to simulate
the production of
(anti)nucleon ($\rm p,\rm\overline p,\rm n, \rm\overline n$) and (anti)hyperon ($\rm\Lambda,\overline\Lambda$)
in Cu+Cu collisions at $\sqrt{s_{\rm{NN}}}=200$~GeV, and to analyze their yields at mid-rapidity ($|\eta|<0.5$). These yields
are compared with experimental data from the STAR collaboration~\cite{agga,agak,star1}
to fix the model parameters. Then, a dynamically constrained phase-space coalescence model(DCPC)~\cite{Yan,chen,chen1} is used to study the production and properties of
light (anti-)nuclei clusters. We expect
that their yields in Cu+Cu collisions may provide some information
about the nature of the  $\rm ^3_{\Lambda}H$ and $\rm ^3_{\overline \Lambda}\overline H$.

\section{Models}
The PACIAE~\cite{sa} is based on PYTHIA 6.4 and designed for various collision types ranging from e+e, p+p, p+A to A+A collisions. In general, PACIAE has four main physics stages consisting of the
parton initiation, parton rescattering, hadronization, and hadron rescattering.
The parton initiation is the first stage in which the nucleus-nucleus collision
is decomposed into the nucleon-nucleon (NN) collisions according to the collision
geometry and NN total cross section. The strings created in the NN collisions will break up into free partons leading to the formation of the deconfined quark gluon matter. After the parton initiation stage,
the decomposed partons rescatter on each other based on the 2 $\rightarrow$ 2 LO-pQCD parton-parton cross
sections~\cite{comb}. Then, the hadronization proceeds throught the Lund string fragmentation model~\cite{sjo} or the phenomenological coalescence model~\cite{sa}. The final stage
is the hadron rescattering process happened between the created hadrons
until the hadronic freeze-out~\cite{sa}.

We calculate the production of light (anti)nuclei and (anti)hypernuclei
with the dynamically constrained phase-space coalescence model (DCPC).
The DCPC model has been studied and used in several collision systems,
such as Au+Au~\cite{chen,chen1,chen2,dong}, Pb+Pb~\cite{chen3}, p+p collisions~\cite{pana}.
We can obtain the yield of a single particle using the following integral

\begin{equation}
Y_1=\int_{H\leqslant E} \frac{d\vec qd\vec p}{h^3},
\end{equation}
where $H$ and $E$ present the Hamiltonian and energy of the particle,
respectively. Then we can also know the yield of N particles calculated by the integral
\begin{equation}
Y_N=\int ...\int_{H\leqslant E} \frac{d\vec q_1d\vec p_1...d\vec
q_Nd\vec p_N}{h^{3N}}. \label{phas}
\end{equation}
This equation, however, has to meet the constraint conditions as follows:
\begin{align}
m_0\leqslant m_{inv}\leqslant m_0+\Delta m;\\
q_{ij}\leqslant D_0(\hspace{0.2cm} i \neq j;j=1,2,...,N).
\end{align}
where
\begin{equation}
m_{inv}=\Big[(\sum_{i=1}^N E_i)^2-(\sum_{i=1}^N p_i)^2\Big]^{1/2}
\end{equation}

$E_i$ and $p_i$ (i=1, 2, ... , N) denote the energies and momenta of particles.
$m_0$ and $\Delta m$ respectively represent the rest mass and the allowed mass uncertainty.
$D_0$ stands for the diameter of (anti)nuclei and (anti)hypernuclei, and $q_{ij}=|\vec q_i-\vec q_j|$
presents the vector distance between particle $i$ and particle $j$. Here,
the diameters are calculated
by $D_0=r_0 \cdot A^{1/3}$, so we choose $r_0=1.4, 1.5, 1.7$ and then get
$D_0=2.02, 2.16, 2.45$ fm for $\rm{^3{He}}$ ($\rm{^3{\overline {He}}}$), $\rm{^3{H}}$ ($\rm{^3{\overline H}}$), $\rm^3_{\Lambda}H$ ($\rm^3_{\overline \Lambda}\overline H$)
in the model, respectively~\cite{e864,sa,ham,yasu}.

\section{Results and discussion}

First the final state particles are produced by the PACIAE
model. In the PACIAE simulations, we assume that the hyperons heavier than $\Lambda$ are already decayed before the creation of hyper-nuclei. The model parameters are selected appropriately to fit the STAR data of p,
$\rm\overline p$, $\Lambda$ and $\overline\Lambda$ in Cu+Cu collisions at $\sqrt{s_{\rm{NN}}}=200 $~GeV for different centrality bins of 0-10\%, 10-30\% and 30-60\%, as shown in Table I, where the yields of particles are calculated with the $|\eta|<0.1$ and $0.4<p_T<1.2$~GeV/c for p and $\rm\overline p$, and $|\eta|<0.5$ and
 $0<p_T<8$~GeV/c for the $\Lambda$ and $\overline\Lambda$. The participants $\langle N_{part}\rangle$ and corresponding experimental data~\cite{agga,agak,star1}
are also presented in Table I. It can be seen from Table I that the yields of particles (p, $\rm\overline p$, $\Lambda$ and $\overline\Lambda$) all decrease rapidly with the increase of centrality, and the PACIAE model results
agree with the STAR data within uncertainties.

Figure 1 shows the transverse momentum distributions of p and $\Lambda$ (open symbols) at the mid-rapidity for
different centralities Cu+Cu collisions at $\sqrt{s_{\rm{NN}}}=200$~GeV measured by PACIAE model.
The solid symbols in this figure are the experimental data taken from Ref.~\cite{agga}. Obviously, one can see from Fig.1 that the transverse momentum distributions of particles (p and $\Lambda$) calculated by PACIAE model are compatible with the STAR data.

\begin{table*}[tbp]
\caption{The participants $\langle N_{part}\rangle$, and yield of particles ($\rm p, \rm \overline p\ ,\Lambda,\overline \Lambda $) in Cu+Cu collisions at $\sqrt{s_{\rm{NN}}}=200$~GeV
for different centrality bins, compared with the STAR data~\cite{agga,agak,star1}.}
\setlength{\tabcolsep}{14.5pt}
\renewcommand{\arraystretch}{1.4}
\begin{tabular}{cccccc}
\hline  \hline
Centrality &$ $&$0-10\%$&$10-30\%$&$30-60\%$ \\ \hline
$\langle N_{part}\rangle$&PACIAE& 103.4 & 67.9 &  28.4\\
$$&STAR&  $99.0\pm1.5$& $64.2\pm1.2$&$29.7\pm0.7$ \\ \hline
p&PACIAE& $5.73\pm0.22$ & $3.55\pm0.23$ & $1.42\pm0.15$  \\
&STAR& $5.56\pm0.02\pm0.19$& $3.35\pm0.01\pm0.12$&  $1.38\pm0.01\pm0.05$ \\
$\rm\overline p $&PACIAE& $4.69\pm0.18$& $2.91\pm0.21$ & $1.14\pm0.11$\\
$$& STAR&  $4.54\pm0.02\pm0.13$& $2.79\pm0.01\pm0.08$& $1.16\pm0.01\pm0.03$ \\
$\Lambda$&PACIAE& $4.24\pm0.13$ & $2.59\pm0.16$ & $1.02\pm0.12$\\
$$&STAR&  4.68$\pm$0.45& 2.67$\pm$0.27&  1.06$\pm$0.11 \\
$\overline \Lambda$&PACIAE& $3.64\pm0.11$ & $2.26\pm0.17$ & $0.91\pm0.10$ \\
$$& STAR&  3.79$\pm$0.37&2.18$\pm$0.23 & 0.88$\pm$0.09 \\
\hline \hline
\\\end{tabular} \label{Npart}
\end{table*}

\begin{figure}[htbp]
\includegraphics[width=0.45\textwidth]{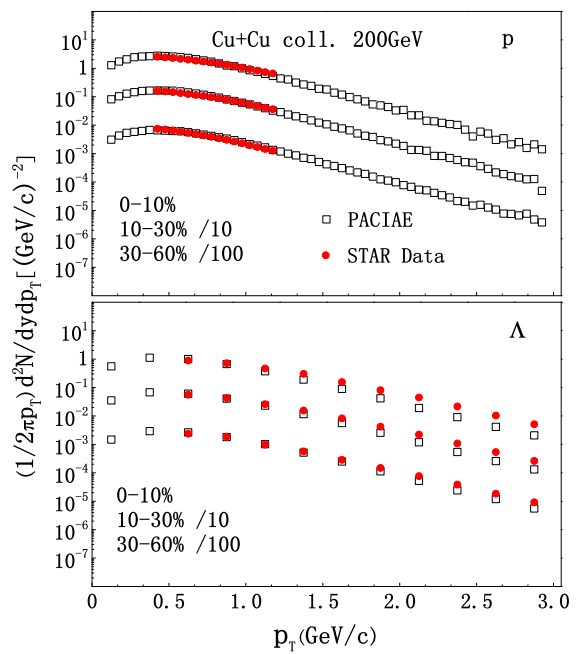}\\
\caption{The transverse momentum spectra of p(upper panel) and $\Lambda$(lower panel)
         in Cu+Cu collisions at $\sqrt{s_{\rm{NN}}}=200$~GeV for different centrality bins of 0-10\%, 10-30\% and 30-60\%. The open squares show the results
         of PACIAE model, and the filled circles show the STAR data~\cite{agga}.}
\label{fig1cucu}
\end{figure}

In the following, the nucleon and hyperon produced within PACIAE are used as input
of the DCPC model. We generate 100 million minimum bias events for Cu+Cu collisions at $\sqrt{s_{\rm{NN}}}=200$~GeV with the PACIAE model. Then we obtain the integrated yields dN/dy for light nuclei and hypernuclei with $|\eta|<0.5$ and $0<p_T<6$~GeV/c for the centrality bins of 0-10\%, 10-30\% and 30-60\%, respectively.
Figure 2 compares their average yield per event at mid-rapidity for central (0-10\%) Cu+Cu collisions at $\sqrt{s_{\rm{NN}}}=200$~GeV with $\Delta m$ varying from 0.3 to 3~MeV. We find that ln(dN/dy) increase linearly with ln$\Delta m$. The behaviors of integrated yields dN/dy of light nuclei and hypernuclei are found to be similar.

\begin{figure}[htbp]
\includegraphics[width=0.5\textwidth]{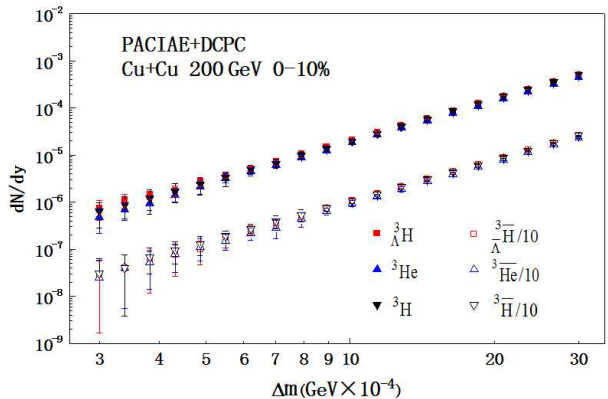}
\caption{Logarithmic distribution of the integrated yield dN/dy of nucleus $\rm ^3_{\Lambda}H$, $\rm{^3{He}}$, $\rm{^3{H}}$ and their anti-nucleus, as a function of the $\Delta m$ with $|\eta|<0.5$ and $0<p_T<6$~GeV/c. For clarity, data of nuclei and anti-nuclei are purposely separated by powers of ten.}
\label{fig2cucu}
\end{figure}

Table II demonstrates the integrated yields dN/dy of $\rm^3_{\Lambda}H$($\rm^3_{\overline \Lambda}\overline H$),
$\rm{^3{H}}$($\rm{^3{\overline H}}$), $\rm{^3{He}}$($\rm{^3{\overline{He}}}$) calculated by {\footnotesize{PACIAE+DCPC}} model
in 0-10\%, 10-30\% and 30-60\% centrality classes Cu+Cu collisions of $\sqrt{s_{\rm{NN}}}=200$~GeV. Here the appropriate parameters $\Delta m$ are fixed to be $\Delta m = 0.79$~MeV for $\rm^3_{\Lambda}H$($\rm^3_{\overline \Lambda}\overline H$), and $\Delta m = 0.89$~MeV for $\rm{^3{H}}$, $\rm{^3{\overline  H}}$, $\rm{^3{He}}$ and $\rm{^3{\overline{He}}}$. The experimental data are taken from STAR Collaboration~\cite{zhou}.
Meanwhile, the {\footnotesize{PACIAE+DCPC}} model results are comparable to that of the experimental data within the current statistical errors.

\begin{table*}[htbp]
\caption{The integrated yields dN/dy of (anti)hypertriton, as well as the (anti)triton and (anti)helium-3 are calculated by {\footnotesize{PACIAE+DCPC}} model in Cu+Cu collisions of $\sqrt{s_{\rm{NN}}}=200$~GeV, compared with experimental data from STAR data~\cite{zhou}.}
\setlength{\tabcolsep}{7.2pt}
\renewcommand{\arraystretch}{1.4}
\begin{tabular}{ccccc}
\hline  \hline
Nucleus type  & STAR &PACIAE$^a($0-10$\%)$ &PACIAE$^a($10-30$\%)$ &PACIAE$^a($30-60$\%)$ \\ \hline
$\rm^3_{\Lambda}H(10^{-5})$  & $-$ & ${1.05\pm0.11}$ & ${0.29\pm0.04}$& ${0.022\pm0.004}$\\
$\rm^3_{\overline \Lambda}\overline H(10^{-5})$  & $-$& ${0.51\pm0.06}$& ${0.14\pm0.02}$& ${0.012\pm0.003}$  \\
$\rm{^3{He}}(10^{-5})$  &$1.29\pm0.22$ &${1.24\pm0.12}$& ${0.37\pm0.05}$& ${0.026\pm0.005}$\\
$\rm{^3{\overline{He}}}(10^{-5})$ &$0.59\pm0.09$ &${0.61\pm0.07}$& ${0.17\pm0.03}$& ${0.014\pm0.003}$\\
$\rm{^3{H}}(10^{-5})$ & $-$ &  ${1.38\pm0.15}$& ${0.35\pm0.05}$& ${0.029\pm0.005}$ \\
$\rm{^3{\overline  H}}(10^{-5})$& $-$ & ${0.74\pm0.10}$& ${0.20\pm0.03}$& ${0.016\pm0.003}$\\
\hline \hline
\multicolumn{5}{l}{$^a$ calculated with $\Delta m=0.79$~MeV for
$\rm^3_{\Lambda}H$, $\rm^3_{\overline \Lambda}\overline H$;
\quad and $\Delta m=0.89$~MeV for
$\rm{^3{He}}$, $\rm{^3{\overline{He}}}$, $\rm{^3{H}}$, $\rm{^3{\overline  H}}$.}\\
\\\end{tabular} \label{ yields}
\end{table*}

In the heavy-ion collisions, production mechanism is generated through
hadron coalescence due to final-state correlations between particles.
The ratios of particle yields can be predicted by the coalescence model,
which have been checked for various particle species.
For instance, the CERN SPS has found that the production of nuclei in
Pb+Pb collisions at $\sqrt{s_{\rm{NN}}}=17.3$~GeV is consistent with
a coalescence picture~\cite{na52}.

Theoretically, the ratios of different (anti)nuclei and (anti)hypernuclei
can be directly related to ratios of hadronic yields in the simple coalescence
framework~\cite{cley,xue}. E.g. if the $\rm^3_{\Lambda}H$ and $\rm^3_{\overline \Lambda}\overline H$
are formed by coalescence of ($\rm p+n+\Lambda$) and ($\rm \overline p+\overline n+\overline\Lambda$),
then the yield ratio of $\rm{^3_{\overline \Lambda}\overline H}/\rm{^3_{\Lambda}H}$ should be
proportional to ($\rm{\overline p}/{p}$)$\times$($\rm {\overline n}/{n}$)$\times$(${\overline\Lambda}/{\Lambda}$),
and the other ratios are the same. The ratios can be written as following:

\begin{equation}
\frac{\rm {^3_{\overline \Lambda}\overline H}}{\rm{^3_{\Lambda}H}}=\\
 \frac{\rm {\overline p \overline n \overline\Lambda}}{\rm{pn\Lambda}}\simeq\\
 (\frac{\rm \overline p}{\rm p})^2\frac{\rm \overline\Lambda}{\Lambda},
\end{equation}
and mixed ratios:
\begin{equation}
\frac{\rm {^3_{\Lambda}H}}{\rm{^3He}}=\\
 \frac{\rm {pn\Lambda}}{\rm{ppn}}\simeq\\
 \frac{\Lambda}{\rm p},
\end{equation}
\begin{equation}
\frac{\rm {^3_{\Lambda}H}}{\rm{^3H}}=\\
 \frac{\rm {pn\Lambda}}{\rm{pnn}}\simeq\\
 \frac{\Lambda}{\rm p},
\end{equation}
\begin{equation}
\frac{\rm {^3_{\overline \Lambda}\overline H}}{\rm{^3\overline{He}}}=\\
 \frac{\rm {\overline p \overline n \overline\Lambda}}{\rm\overline{ppn}}\simeq\\
 \frac{\rm \overline\Lambda}{\rm\overline p},
\end{equation}
\begin{equation}
\frac{\rm {^3_{\overline \Lambda}\overline H}}{\rm{^3\overline{H}}}=\\
 \frac{\rm {\overline p \overline n \overline\Lambda}}{\rm\overline{pnn}}\simeq\\
 \frac{\rm \overline\Lambda}{\rm\overline p}.
\end{equation}

\begin{figure}[htbp]
\includegraphics[width=0.5\textwidth]{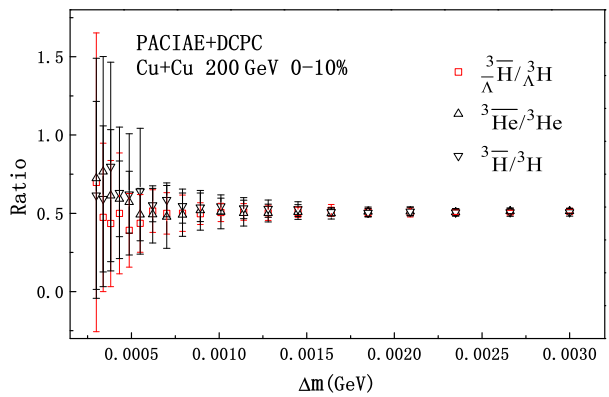}
\caption{The distribution of ratios ${\rm{^3_{\overline \Lambda}}\overline H}/{\rm{^3_{\Lambda}H}}$, $\rm{{^3{\overline He}}/{^3{He}}}$ and $\rm{{^3{\overline H}}/{^3{H}}}$ versus the mass uncertainty
 ($\Delta m$) in Cu+Cu collisions at $\sqrt{s_{\rm{NN}}}=$200~GeV($|\eta|<0.5$ and $0<p_T<6$~GeV/c). Error bars are statistical only.}
\label{fig3cucu}
\end{figure}

The ratios $\rm{{^3_{\overline \Lambda}\overline H}/{^3_{\Lambda}H}}$,
$\rm{{^3{\overline{He}}}/{^3{He}}}$ and $\rm{{^3{\overline H}}/{^3H}}$ calculated by {\footnotesize{PACIAE+DCPC}} model with $|\eta|<0.5$ and $0<p_T<6$~GeV/c, as a function of the $\Delta m$, as shown in Fig.3. It is clear that the ratios are close to 0.5 for almost all the mass uncertainty ($\Delta m$) range. The ratios of anti-nuclei to nuclei are independent of $\Delta m$, although their integrated yield dN/dy has a strong dependence on $\Delta m$ as shown in Fig.2.

\begin{table*}[htbp]
\caption{The header of the table presents the (anti)nucleus ratios from {\footnotesize{PACIAE+DCPC}} model in Cu+Cu collisions at $\sqrt{s_{\rm{NN}}}=200$~GeV for three different centrality classes. The top section of the table shows the three ratios of anti-nucleus to nucleus, followed by the mixed ratios of (anti)nucleus to (anti)nucleus. And the ratios between proton, antiproton, hyperon, anti-hyperon are shown at the bottom. In the top two sections, STAR data are taken from Au+Au collisions at $\sqrt{s_{\rm{NN}}}=200$~GeV~\cite{abel} and in the bottom, STAR data are taken from Cu+Cu collisions at $\sqrt{s_{\rm{NN}}}=200$~GeV~\cite{agga,agak}.}
\setlength{\tabcolsep}{7.2pt}
\renewcommand{\arraystretch}{1.4}
\begin{tabular}{cccccc}
\hline  \hline
Ratio & STAR &PACIAE(0-10\%)&PACIAE(10-30\%)&PACIAE(30-60\%)\\\hline
$\rm{{^3_{\overline \Lambda}\overline H}/{^3_{\Lambda}H}}$&${0.49\pm0.18\pm0.07}$&${0.50\pm0.06}$&${0.49\pm0.07}$ &${0.54\pm0.12}$ \\
$\rm{{^3{\overline{He}}}/{^3{He}}}$  &${0.45\pm0.02\pm0.04}$ &${0.51\pm0.06}$&${0.46\pm0.06}$&${0.53\pm0.11}$ \\
$\rm{{^3{\overline H}}/{^3H}}$ & $-$ &${0.54\pm0.07}$ &${0.58\pm0.09}$&${0.56\pm0.12}$ \\
\hline
$\rm{{^3_{\overline \Lambda}\overline H}/{^3{\overline{He}}}}$&${0.89\pm0.28\pm0.13}$&${0.84\pm0.10}$ &${0.83\pm0.12}$&${0.86\pm0.21}$ \\
$\rm{{^3_{\Lambda}H}/{^3{He}}}$ &${0.82\pm0.16\pm0.12}$&${0.85\pm0.08}$ &${0.80\pm0.12}$&${0.85\pm0.20}$ \\
$\rm{{^3_{\overline \Lambda}\overline H}/{^3{\overline H}}}$&$-$&${0.70\pm0.09}$&${0.71\pm0.11}$&${0.76\pm0.19}$ \\
$\rm{{^3_{\Lambda}H}/{^3{H}}}$&$-$&${0.77\pm0.11}$&${0.82\pm0.14}$ &${0.75\pm0.18}$ \\\hline
$\rm{\overline p /p}$& 0.80$\pm$0.04& $0.81\pm0.05$   & $0.82\pm0.08$  &  $0.80\pm0.11$ \\
$\overline \Lambda /\Lambda$& 0.82$\pm$0.12& $0.86\pm0.04$   & $0.87\pm0.09$  &  $0.88\pm0.15$   \\
$\Lambda /p$& 0.84$\pm$0.09& $0.74\pm0.04$   & $0.73\pm0.07$  &  $0.72\pm0.12$ \\
$\overline \Lambda /\overline p$& 0.83$\pm$0.08& $0.78\pm0.05$   & $0.78\pm0.08$  &  $0.80\pm0.12$   \\
\hline \hline
\\\end{tabular} \label{Particle ratios}
\end{table*}

\begin{figure}[htbp]
\includegraphics[width=0.5\textwidth]{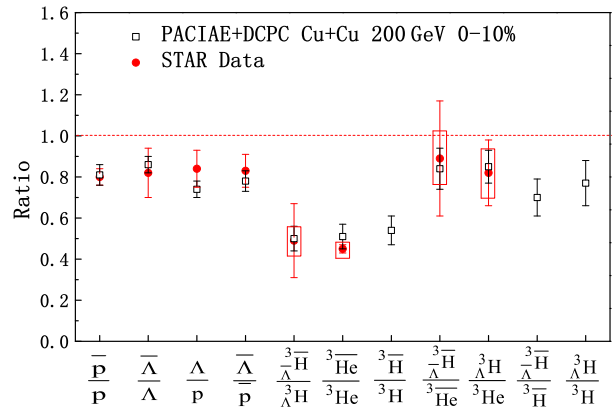}
\caption{The ratios ($\overline p/p$, $\rm{\overline \Lambda}/{\Lambda}$, $\rm{{^3_{\overline \Lambda}\overline H}/{^3_{\Lambda}H}}$, $\rm{{^3{\overline{He}}}/{^3{He}}}$, $\rm{{^3{\overline H}}/{^3H}}$) and ($ p/{\Lambda}$, $\rm{\overline \Lambda}/\overline p$, $\rm{{^3_{\Lambda}H}/{^3{He}}}$,
$\rm{{^3_{\overline \Lambda}\overline H}/{^3{\overline{He}}}}$,
$\rm{{^3_{\Lambda}H}/{^3{H}}}$, $\rm{{^3_{\overline \Lambda}\overline H}/{^3{\overline H}}}$ )
 determined by the {\footnotesize{PACIAE+DCPC}}
  model analysis (open squares) for matter and anti-matter compared with STAR results (filled circles)~\cite{abel}. Here, statistical and systematic errors are represented by error bars and error boxes, respectively.}
\label{fig4cucu}
\end{figure}

Table III represents ratios of anti-particles to corresponding particles ($\rm \overline p/p$, $\rm{\overline \Lambda}/{\Lambda}$, $\rm{{^3_{\overline \Lambda}\overline H}/{^3_{\Lambda}H}}$,
$\rm{{^3{\overline{He}}}/{^3{He}}}$, $\rm{{^3{\overline H}}/{^3H}}$) and their mixed ratios ($ \rm {\Lambda}/p$, $\rm{\overline \Lambda}/\overline p$, $\rm{{^3_{\Lambda}H}/{^3{He}}}$,
$\rm{{^3_{\overline \Lambda}\overline H}/{^3{\overline{He}}}}$,
$\rm{{^3_{\Lambda}H}/{^3{H}}}$, $\rm{{^3_{\overline \Lambda}\overline H}/{^3{\overline H}}}$ )
for the centrality bins of 0-10\%, 10-30\% and 30-60\% in Cu+Cu collisions at $\sqrt{s_{\rm{NN}}}=200$~GeV, where, $\rm^3_{\Lambda}H$, $\rm^3_{\overline \Lambda}\overline H$ are calculated with $\Delta m=0.79$~MeV and $\rm{^3{He}}$, $\rm{^3{\overline{He}}}$, $\rm{^3{H}}$, $\rm{^3{\overline H}}$ with $\Delta m=0.89$~MeV. The result with centrality bin of 0-10\% is drawn in Fig.4.
We can see from Table III, the yield ratios of $\rm \overline p/p$, $\rm{\overline \Lambda}/{\Lambda}$, $\rm{{^3_{\overline \Lambda}\overline H}/{^3_{\Lambda}H}}$,
$\rm{{^3{\overline{He}}}/{^3{He}}}$, $\rm{{^3{\overline H}}/{^3H}}$ and their mixed ratios $\rm {\Lambda}/p$, $\rm{\overline \Lambda}/\overline p$, $\rm{{^3_{\Lambda}H}/{^3{He}}}$,
$\rm{{^3_{\overline \Lambda}\overline H}/{^3{\overline{He}}}}$,
$\rm{{^3_{\Lambda}H}/{^3{H}}}$, $\rm{{^3_{\overline \Lambda}\overline H}/{^3{\overline H}}}$ are all independent of the centrality bin, although their yields decrease rapidly with the centrality as shown in Table I and Table II.
The ratio of the (anti)hypertriton to (anti)nuclei ($\rm{{^3_{\Lambda}H}/{^3{He}}}$,
$\rm{{^3_{\overline \Lambda}\overline H}/{^3{\overline{He}}}}$,
$\rm{{^3_{\Lambda}H}/{^3{H}}}$, $\rm{{^3_{\overline \Lambda}\overline H}/{^3{\overline H}}}$) are less than 1, which means that the yield of (anti)hypertriton is less than that of (anti)nuclei.

Table III and Fig.4 show that the ratios of the (anti)hypernuclei to (anti)nuclei ($\rm{{^3_{\Lambda}H}/{^3{He}}}$,$\rm{{^3_{\overline \Lambda}\overline H}/{^3{\overline{He}}}}$,
$\rm{{^3_{\Lambda}H}/{^3{H}}}$, $\rm{{^3_{\overline \Lambda}\overline H}/{^3{\overline H}}}$ ), as Eqs.(6)-(10) predict, are compatible to the ratio of hyperon to proton within uncertainties.
The simulation results in our model are found to be in agreement with the experimental data from STAR (0-80\% centrality) Au+Au collisions at $\sqrt{s_{\rm{NN}}}=200$~GeV~\cite{abel,agga,agak}. Furthermore, the model predictions of $\rm{{^3{\overline H}}/{^3H}}$,
$\rm{{^3_{\Lambda}H}/{^3{H}}}$ and $\rm{{^3_{\overline \Lambda}\overline H}/{^3{\overline H}}}$ are also presented in Fig.4.

\begin{figure}[htbp]
\includegraphics[width=0.45\textwidth]{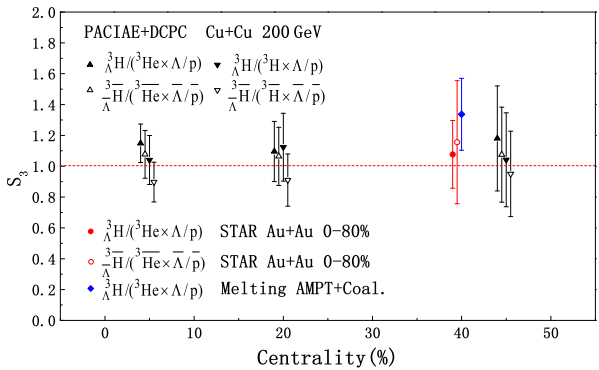}
\caption{The $S_3$(${\overline S_3}$) ratio as a function of the centrality in Cu+Cu collisions at $\sqrt{s_{\rm{NN}}}=200$~GeV. For comparison, the data from STAR Au+Au 0-80\% collisions at $\sqrt{s_{\rm{NN}}}=200$~GeV~\cite{abel} and minimum bias Au+Au collisions from melting AMPT plus coalescence model calculation~\cite{zhang} are shown here. Statistical uncertainties are represented by bars.}
\label{fig5cucu}
\end{figure}

The strangeness population factor for light nuclei $S_3=\rm{{^3_{\Lambda}H}/({^3{He}}\times \Lambda/p)}$ in a naive coalescence model, should be a value near unity, as the Ref.~\cite{e864} shows. The ratio is sensitive to the local baryon-strangeness correlation~\cite{koch,maju,mcheng,adams,braun1,gavai}, hence it can provide a possible chance to study the nature of matter created in the collision~\cite{sato,zhang}. The $S_3(\overline S_3)$ values obtained by the {\footnotesize{PACIAE+DCPC}} model for different centrality bins of 0-10\%, 10-30\% and 30-60\% are shown in Fig.5 and Table IV. These values are compared with theoretical model~\cite{zhang} and the data from STAR~\cite{abel}.

The models used for the comparison are the string melting of the AMPT (A Multi-Phase Transport Model for Relativistic Heavy Ion Collisions)~\cite{AMPT} plus coalescence described in Ref.~\cite{zhang}. The present results in Cu+Cu collisions at $\sqrt{s_{\rm{NN}}}=200$~GeV show the values of $S_3(\overline S_3)$ close to unity, indicating that the phase-space populations for strange and light quarks are similar implying the high-temperature deconfined quarks can be formed in Cu+Cu collisions. These results are also consistent with the STAR experiment~\cite{abel} and the AMPT with string melting plus coalescence model in Au+Au collisions at $\sqrt{s_{\rm{NN}}}=200$~GeV within uncertainties.

\begin{table*}[htbp]
\caption{$S_3$ ratio for nucleus and antinucleus measured by {\footnotesize{PACIAE+DCPC}} model in Cu+Cu collisions of $\sqrt{s_{\rm{NN}}}=200$~GeV for three different centrality classes. The data for comparison are taken from STAR Au+Au 0-80\% collisions at 200~GeV~\cite{abel} and  minimum bias Au+Au collisions melting AMPT plus coalescence model calculation~\cite{zhang}.}
\setlength{\tabcolsep}{7.2pt}
\renewcommand{\arraystretch}{1.4}
\begin{tabular}{cccccc}
\hline  \hline
&Centrality & $\rm{{^3_{\Lambda}H}/({^3{He}\times\Lambda/p)}}$  &$\rm{{^3_{\overline \Lambda}\overline H}/({^3{\overline{He}}\times{\overline\Lambda}/\overline p)}}$   & $\rm{{^3_{\Lambda}H}/({^3{H}}\times\Lambda/p)}$  &$\rm{{^3_{\overline \Lambda}\overline H}/({^3{\overline H}}\times{\overline\Lambda}/\overline p)}$ \\ \hline
&0-10\%  & ${1.15\pm0.12}$ & ${1.08\pm0.15}$ & ${1.04\pm0.16}$& ${0.90\pm0.13}$\\
{\footnotesize{PACIAE+DCPC}}&10-30\% & ${1.10\pm0.20}$ & ${1.06\pm0.19}$& ${1.12\pm0.22}$& ${0.91\pm0.17}$  \\
&30-60\% &$1.18\pm0.34$ &${1.08\pm0.31}$& ${1.04\pm0.30}$& ${0.95\pm0.27}$\\
\hline
STAR&0-80\% &$1.08\pm0.22$ &${1.16\pm0.40}$&$-$&$-$\\
Melting AMPT+Coal.&0-80\%  &$1.34\pm0.23$ &$-$&$-$&$-$\\
\hline \hline
\\\end{tabular} \label{ S3}
\end{table*}

\section{Conclusion}
In the paper, we have used the {\footnotesize{PACIAE}} model to simulate Cu+Cu collisions at a center-of-mass energy $\sqrt{s_{\rm{NN}}}=200$~GeV. The obtained yields of final state particles $\rm p$, $\rm {\overline p}$, n, $\rm\overline n$, ${\Lambda}$ and ${\overline \Lambda}$ within {\footnotesize{PACIAE}} model are in very good agreement with experimental data from STAR. The $\rm p$, $\rm {\overline p}$, n, $\rm\overline n$, ${\Lambda}$ and ${\overline \Lambda}$ are used as input for the {\footnotesize DCPC} model to construct $\rm{^3_{\Lambda}H}$, $\rm{^3_{\overline \Lambda}\overline H}$, $\rm{^3{He}}$, $\rm{^3{\overline{He}}}$, $\rm{{^3{H}}}$ and $\rm{{^3{\overline H}}}$ clusters through coalescence.
A comparison between $\rm{^3_{\Lambda}H}$, $\rm{^3_{\overline \Lambda}\overline H}$ and $\rm{^3{He}}$, $\rm{^3{\overline{He}}}$, $\rm{{^3{H}}}$, $\rm{{^3{\overline H}}}$ in Cu+Cu collisions at $\sqrt{s_{\rm{NN}}}=200$~GeV for three centrality classes(0-10\%, 10-30\% and 30-60\%) has been presented.

The results show that yields of $\rm{^3_{\Lambda}H}$, $\rm{^3_{\overline \Lambda}\overline H}$ $\rm{^3{He}}$, $\rm{^3{\overline{He}}}$, $\rm{{^3{H}}}$ and $\rm{{^3{\overline H}}}$ decrease rapidly with the increase of centrality bins, but their yield ratios are independent of centrality, which is consistent with the STAR experimental data.  The ratios of the (anti)hypertriton to (anti)nuclei($\rm{{^3_{\Lambda}H}/{^3{He}}}$,
$\rm{{^3_{\overline \Lambda}\overline H}/{^3{\overline{He}}}}$,
$\rm{{^3_{\Lambda}H}/{^3{H}}}$, $\rm{{^3_{\overline \Lambda}\overline H}/{^3{\overline H}}}$) are less than 1, which means that the yield of (anti)hyper-triton is less than that of (anti)nuclei.
The strangeness population factor $S_3=\rm{{^3_{\Lambda}H}/({^3{He}}\times \Lambda/p)}$ for matter and anti-matter with the tritium and helium-3 is measured to be close to unity, which is compatible with the STAR data and the AMPT with string melting plus coalescence model, suggesting that the phase-space population of strange is similar with the ones of light quarks and support the generation of high-temperature matter of deconfined quarks. The yields and yield ratios of $\rm ^3_{\Lambda}H$($\rm ^3_{\overline \Lambda}\overline H$), $\rm ^3H(^3\overline H)$ and $\rm ^3He(^3\overline {He})$ in Cu+Cu collisions at $\sqrt{s_{\rm{NN}}}=200$~GeV, are first predicted with the theoretical model.

\begin{center} {ACKNOWLEDGMENT} \end{center}
The financial support from NSFC (11475149) is acknowledged, and supported by the high-performance computing
platform of China University of Geosciences. The authors thank PH.D. Huan Chen and Wei Dai for helpful discussions.

\end{document}